\documentclass[aps,pre,superscriptaddress,twocolumn,showpacs]{revtex4}

\usepackage{graphicx}
\usepackage[latin1]{inputenc}

\begin{document}

\title{Hydrodynamic lift of vesicles under shear flow in microgravity}

\author{Natacha Callens}
\author{Christophe Minetti}
\affiliation{Microgravity Research Center, Universit\'e Libre de Bruxelles, 50 Av. F. Roosevelt, B-1050 Brussels, Belgium}
\author{Gwennou Coupier}
\author{Maud-Alix Mader}
\affiliation{Laboratoire de Spectrom\'etrie Physique, CNRS Universit\'e J. Fourier - Grenoble I, BP 87, 38402 Saint Martin d'H\`eres, France}
\author{Frank Dubois}
\affiliation{Microgravity Research Center, Universit\'e Libre de Bruxelles, 50 Av. F. Roosevelt, B-1050 Brussels, Belgium}
\author{Chaouqi Misbah}
\author{Thomas Podgorski}\email{thomas.podgorski@ujf-grenoble.fr}
\affiliation{Laboratoire de Spectrom\'etrie Physique, CNRS Universit\'e J. Fourier - Grenoble I, BP 87, 38402 Saint Martin d'H\`eres, France}

\date{April 9, 2008}

\begin{abstract}
The dynamics of a vesicle suspension in a shear flow between parallel plates has been investigated under microgravity conditions, where vesicles are only submitted to hydrodynamic effects such as lift forces due to the presence of walls and drag forces. The temporal evolution of the spatial distribution of the vesicles has been recorded thanks to digital holographic microscopy, during parabolic flights and under normal gravity conditions. The collected data demonstrates that vesicles are pushed away from the walls with a lift velocity proportional to $\dot{\gamma} R^3/z^2$ where $\dot{\gamma}$ is the shear rate, $R$ the vesicle radius and $z$ its distance from the wall. This scaling as well as the dependence of the lift velocity upon vesicle aspect ratio are consistent with theoretical predictions by Olla [J. Phys. II France {\bf 7}, 1533--1540 (1997)].
\end{abstract}


\maketitle

\section{Introduction}

Vesicles are closed lipid membranes suspended in an aqueous
solution. They can be considered as a simple mechanical model for living cells, especially simple ones like red blood cells. Their behaviour under shear flow has been the subject of several theoretical \cite{KS,Beaucourt04,Gompper04,Misbah06} and experimental \cite{Haas97,Abkarian,Mader06,Steinberg05,Steinberg06} studies. 
When the viscosity ratio between the inner and outer fluids is small   \cite{KS}, vesicles follow a tank treading dynamics in which the orientation of the main axis of the vesicle is constant and the membrane undergoes a tank treading motion. When they are sheared close to a wall, a lift force appears which pushes vesicles away from the wall \cite{Olla97, Cantat99, Seifert99, Sukumaran01, Abkarian}. If gravity is perpendicular to the wall, vesicles reach a stationary altitude where the lift force is balanced by vesicle weight. Because of the rapid decrease of the lift force when the distance increases, the equilibrium distance under gravity is usually of order of the vesicle size \cite{Abkarian}.

The unbinding and lift of vesicles from a wall have been investigated theoretically and numerically by several authors. Cantat and Misbah \cite{Cantat99} and Seifert \cite{Seifert99} studied the dynamical unbinding of adhering vesicles and the lift force when the distance between vesicles and the wall is small. Olla \cite{Olla97} made a theoretical prediction of the drift velocity of a neutrally buoyant ellipsoidal vesicle when the distance to the wall is large compared to vesicle size. He found:
\begin{equation}
\frac{d z}{d t} = U(\lambda, r_2, r_3) \dot{\gamma} \frac{R^3}{z^2},
\label{eq-olla}
\end{equation}
where $\dot{\gamma}$ is the shear rate, $R=(a_1 a_2 a_3)^{1/3}$ the vesicle's effective radius, $a_1$, $a_2$, $a_3$ the ellipsoid's semi-axes and $z$ the distance between the vesicle center and the wall. $U$ is a dimensionless drift velocity which depends on the viscosity ratio $\lambda$ and the axis ratios $r_2=a_2/a_1$ and $r_3=a_3/a_1$ of the ellipsoid.
Sukumaran and Seifert \cite{Sukumaran01} made a numerical study of the lift in the vicinity of a wall and found a similar form for the lift velocity.

An experimental investigation of the unbinding and lift of vesicles under gravity was performed by Abkarian et al. \cite{Abkarian}. By measuring the equilibrium height of vesicles of known density, they suggested the following empirical law for the lift force:
\begin{equation}
F_l \sim \eta \dot{\gamma} \frac{R^3}{h},
\end{equation}
where $h$ is the distance between the vesicle membrane and the wall ($h\simeq z-R$).
Assuming a Stokes-like drag force $F_d \sim \eta (dz/dt) R$, this would lead to a lift velocity for neutrally buoyant vesicles that scales like $\dot{\gamma} {R^2}/{h}$. This markedly differs from eq. \ref{eq-olla}.
However this experimental study is limited to short distances from the wall where lubrication effects are strong, and in a situation where gravity may also affect vesicle shape. This may result in very different scaling laws. To our knowledge no experimental study deals with the drift of vesicles at larger distances from a wall in a shear flow.

The ESA project BIOMICS (BIOMImetic and Cellular Systems) is dedicated to the investigation of the dynamics of a suspension of vesicles under shear flow in microgravity conditions. Within the framework of this project several parabolic flight campaigns (ESA PF 43, CNES VP 59 and CNES VP 65) on board the Airbus A300-Zero G from Novespace (Merignac, France) were performed and an experiment will be part of the MASER 11 sounding rocket payload.

In this paper we present experiments, realized during the VP65 parabolic flight campaign, aiming at studying the lift of vesicles initially close to a wall under microgravity conditions where only hydrodynamic forces are present. 
A digital holographic microscope (DHM), working in transmission with a partial spatial coherent source, allows a 3D visualisation of the evolution of vesicle positions in the field of view during microgravity \cite{DHM}. 

The results show a drift of vesicles from the wall, at distances greater than vesicle size, which is compatible with lift forces given by eq. \ref{eq-olla}. It underlines that the spatial distribution of vesicles across the channel thickness is strongly influenced by the shear flow and vesicle deformability.

\section{Experimental setup}

\begin{figure}
\centering
\includegraphics[width=8cm]{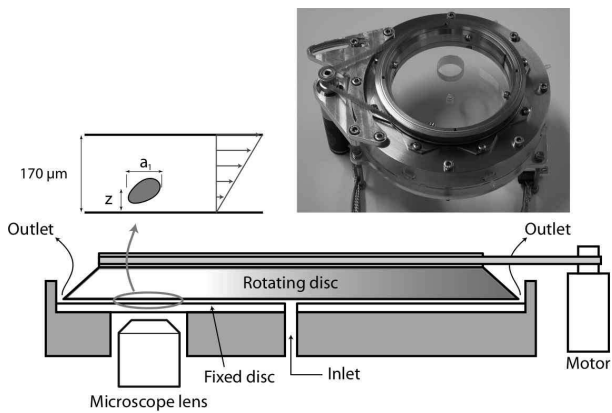} 
\caption{Sketch and picture of the shear flow chamber developed by SSC (color online). \label{fig_chamber2}}
\end{figure}

The experiments were performed in a Couette shear flow chamber
designed and manufactured by the Swedish Space Corporation (SSC) for the MASER 11 sounding rocket flight foreseen for May 2008. As shown on Fig. \ref{fig_chamber2}, it is made of two parallel glass discs, with a diameter of 10 cm and a gap of 170 $\mu$m. The bottom disc is fixed and the top one can be rotated at constant speed by a computer controlled motor via a belt and a reducer, allowing shear rates between 0.5 and 50 s$^{-1}$.
An inlet at the center of the bottom plate and several outlets at the periphery allow filling with experimental fluids, vesicle injection and rinsing. Vesicle suspensions and extra carrier fluid are contained in syringes connected to the inlet via flexible tubes. A KDS syringe pump is used for the injections, and a soft pouch collects the excess and waste fluids. 

\begin{figure}
\begin{center}
\includegraphics[width=6cm]{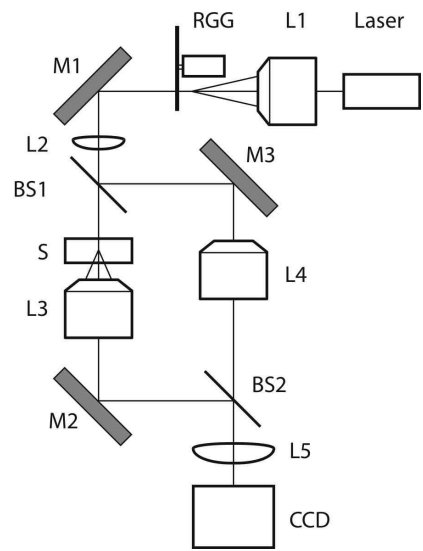} 
\end{center}
\caption{Optical setup of the digital holographic microscope. L1: focusing lens; RGG: rotating ground glass for spatial coherence reduction; L2: collimating lens; L3, L4: identical microscope lens (x 20); L5: refocusing lens; CCD: charge-coupled device camera; M1-M3: mirrors; BS1, BS2: beam splitters; S: shear flow chamber\label{fig_DHM}}
\end{figure}

The observation system used to characterize the evolution and dynamics of the suspension is a digital holographic microscope working with a partially coherent illumination designed and manufactured by MRC \cite{DHM}. Digital holography is an emerging technology that provides 3D imaging of microscopic samples. The main advantage of digital holography is its capability to record rapidly the 3D information without scanning along the optical direction by changing the focus as with usual microscopy or confocal microscopy. Therefore it provides the capability to investigate highly dynamic phenomena. This is a decisive advantage for 3D flow analysis. The system takes benefit of a source of partial spatial coherence, created with a laser beam incident on a rotating ground glass that reduces the amount of coherent artifact noise and the multiple reflection perturbations \cite{reflection}. The optical system is described by the Fig. \ref{fig_DHM}. It is a Mach-Zehnder interferometer in microscope configuration. The shear flow chamber is placed in one arm of the interferometer in front of a microscope lens. The direction of observation is perpendicular to the plates, approximately 3 cm from the axis of rotation. The microscope lens, coupled with the refocusing lens, produces the image of one plane of the shear flow chamber thickness on the CCD. The second arm of the interferometer constitutes the reference beam and is also incident on the CCD where it is interfering with the object beam. 

By recording the holographic data resulting from the interference pattern, the digital holographic microscope provides the capability to refocus numerically vesicles that are out of focus with respect to the focus plane of the optical imaging subsystem. The holograms are recorded by a CCD camera and the digital holographic process, which consists of the refocusing in depth, is achieved by computation. 
The digital holographic reconstruction increases, without scanning, the depth of investigation by a factor of 100 in comparison with classical optical microscopy. The phase and intensity information are extracted from every hologram by the Fourier method  \cite{Takeda99, Fourier} and the complex amplitude is computed and used to refocus it in any plane inside the shear flow chamber thickness  \cite{bestf}.
The field of view of the digital holographic microscope is (400 $\mu$m)$^3$. The image acquisition system is able to record 24 holograms/second. As the vesicles are transparent objects, when they are focus, it is necessary to take benefit of the phase information, to locate them in the experimental volume. Digital holographic reconstruction coupled to specific algorithmic techniques for object detection  \cite{bestf, border} permits to obtain the size, projected shape in the XY plane, 3D position, and an estimation of the mean velocity of each flowing vesicle.

\section{Sample preparation and experimental procedures}

Vesicles were prepared by the electroformation method  \cite{angelova92}, from lipids (1,2 - Dioleoyl-sn-glycero-3-phosphocholine (DOPC) from Sigma-Aldrich (P-6354)) deposited on ITO glass plates and an electroformation solution made of water, glycerol and sucrose (S-7903 from Sigma). Each electroformation provides a polydisperse sample with vesicle diameter ranging from 5 to 100 $\mu$m (median size between 15 and 25 $\mu$m), a volume fraction of a few percent and a total volume of about 2 ml.

Samples were then diluted into a slightly hyperosmotic exterior solution made of water, glycerol and glucose (G-7528 from Sigma), which leads to a partial deflation of vesicles, a difference of refractive index necessary for visualization in the digital holographic microscope and a density contrast responsible for sedimentation under gravity. The properties of internal and external solutions are summarized in table \ref{tabsol}. The internal and external viscosities are nearly identical (viscosity ratio $\lambda=0.99$), and the density difference is $0.01$ g/l, leading to sedimentation velocities under normal gravity of about 20 $\mu$m/s for vesicles with a radius of 10 $\mu$m.

\begin{table}
\begin{center}
\begin{tabular}{lccc}
\hline
Solution & Content & Viscosity & Density \\
\hline
interior & 10\% glycerol  & $1.51$ cP & $1.045$ g/l \\
         & + 200 mM sucrose &         &             \\        
exterior & 10\% glycerol  & $1.55$ cP & $1.035$ g/l \\
         & + 230 mM glucose &         &             \\        

\hline
\end{tabular}
\end{center}
\caption{Composition and physical properties of sample internal and external solutions \label{tabsol}}
\end{table}

The experimental procedure during flight is as follows: 
A parabolic flight consists in 31 parabolas, each providing 22 seconds of microgravity (with a tolerance of $\pm 5. 10^{-2}$ g). Between parabolas (about 2.5 min), phases of normal and hyper gravity (1.5 to 1.8 g), during which the rotation of the shear flow chamber is stopped, allow a rapid sedimentation of the vesicles and provide a simple and reproducible initial condition with all vesicles lying on the bottom (fixed) disc. At the beginning of microgravity phases, the rotation of the chamber starts with shear rates between 0.5 and 50 s$^{-1}$ and images are recorded at 24 holograms/s through the DHM.

\section{Data processing}

\begin{figure}
\begin{center}
\includegraphics[width=8cm]{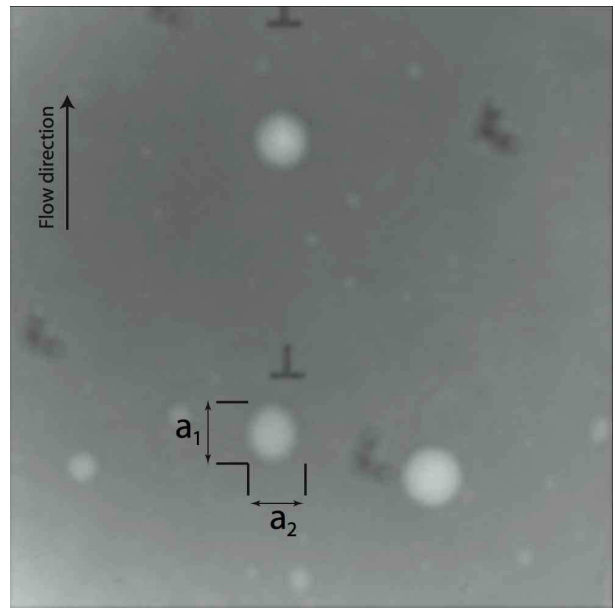} 
\end{center}
\caption{Example of a phase map extracted from an hologram recorded in microgravity. White objects are vesicles, dark shapes are markers on glass plates (T-shapes on the bottom and V-shapes on the top plate).\label{fig_phase}}
\end{figure}

To obtain reliable statistics, algorithms and techniques have been developed in order to get automatic tools for the data processing. Indeed, for each parabola, 600 holograms are recorded leading to a huge amount of data. To obtain the 3D behaviour of vesicles flowing in the field of view, each hologram is reconstructed on the whole shear flow chamber thickness. From this set of reconstructed planes, the x and y position of each vesicle is obtained by phase contrast analysis. The z position (over the chamber thickness) is then determined using a specific metric  \cite{bestf}. Knowing the position of each detected object, their sizes and shapes (projected on the XY plane) are obtained by segmenting the phase map in the respective focus plane (Fig. \ref{fig_phase}).

Collecting the 3D behaviour of many vesicles passing in the field of view during microgravity time allows to gather statistical data about the evolution of the distribution of vesicles in the chamber thickness - due to lift forces and hydrodynamic interactions - as a function of time, as well as the influence of parameters such as vesicle size and deflation.
Between 5 and 20 vesicles with a diameter greater than 10 $\mu$m are detected on each image. We get the spatial distribution and shape of 
several thousand vesicles for each microgravity period.

Further data processing consists in computing each detected vesicle's effective radius $R=(a_1 a_2^2)^{(1/3)}/2$ where $a_1$ and $a_2$ are the measured long and short axis of the projected vesicle shape (fitted by an ellipse) and its aspect ratio $a_1/a_2$. Note that these are apparent (projected) parameters. A vesicle's real radius and aspect ratio are slightly bigger due to the inclination of tank-treading vesicles in shear flow \cite{KS, Olla97, Abkarian} and can be recovered by assuming a prolate ellipsoidal shape (where the two small axes are identical) and using the theoretical relation between vesicle deflation and inclination from \cite{KS, Olla97}. Then, filtering on size $R$, aspect ratio $a_1/a_2$ and averaging of results over a prescribed number of vesicles or consecutive frames can be performed.

\section{Results and discussion}

\begin{figure}
\centering
\includegraphics[width=8.5cm]{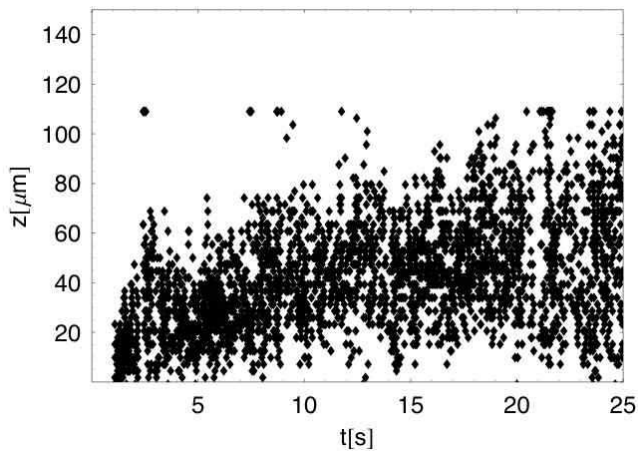} 
\caption{Distribution of vesicle distances from the wall during microgravity for a shear rate of 50 s$^{-1}$, $5<R<50~\mu$m and $1<a_1/a_2<2$ \label{fig_zraw}}
\end{figure}

As the shear flow starts under microgravity, polydisperse vesicles are lifted away from the bottom wall with drift velocities which depend on vesicle size and shape. An example is show on Fig. \ref{fig_zraw} where the vesicle cloud is seen to drift towards the shear flow chamber center.

The large dispersion on this graph is due to several factors: the main factor is the polydispersity of the sample (both in size and aspect ratio): in a typical sample, detected vesicle radii range from 3 to 50 $\mu$m and aspect ratios (long axis over short axis) between 1 and 2. This leads to very different lift velocities. Hydrodynamic interactions between vesicles also lead to a dispersion of objects: as the suspension is sheared, vesicles move around each other and repel. Finally errors from the optical technique and data processing can be estimated to $\pm 3 \mu$m.

For more quantitative results, it is therefore necessary to find a proper scaling and/or filter data to make results less sensitive to polydispersity, and to average results over several consecutive frames (or several detected objects with similar properties) in order to reduce the effect of noise in the data.

\begin{figure}
\centering
\includegraphics[width=8.5cm]{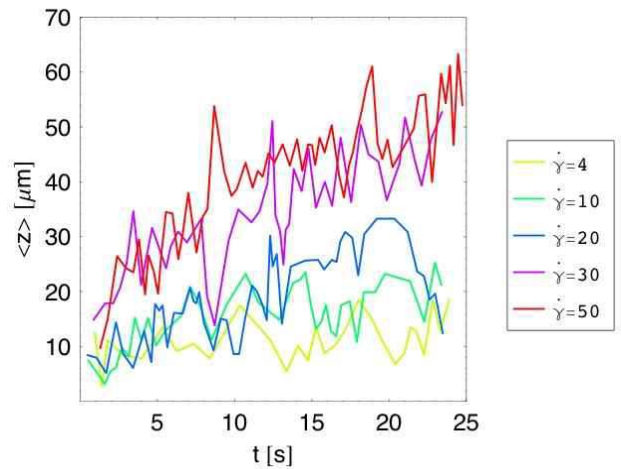} 
\caption{Average vesicle distance from the wall vs. time under microgravity conditions for different shear rates (filtering parameters: $5<R<10~\mu$m, $1.1<a_1/a_2<1.25$ ; averaging over 20 detections). \label{fig_zavg}}
\end{figure}

Figure \ref{fig_zavg}
shows the result of such a treatment for different experiments at different shear rates, where only vesicles in narrow ranges of sizes and aspect ratios were kept. The effect of the shear rate is clearly visible: vesicles drift much faster at higher shear rates $\dot{\gamma}$. Note that no saturation is observed, indicating that vesicles are still quite far from the mid plane of the shear flow chamber (located at $z=85~\mu$m) and that the lift force is still significant at a distance from the wall of order a few vesicle diameters.

By integrating Olla's result (eq. \ref{eq-olla}), one gets the following scaling law:
\begin{equation}
(\frac{z}{R})^3 = 3 U \dot{\gamma} t 
\label{scaling}
\end{equation}

\begin{figure}
\centering
\includegraphics[width=8.5cm]{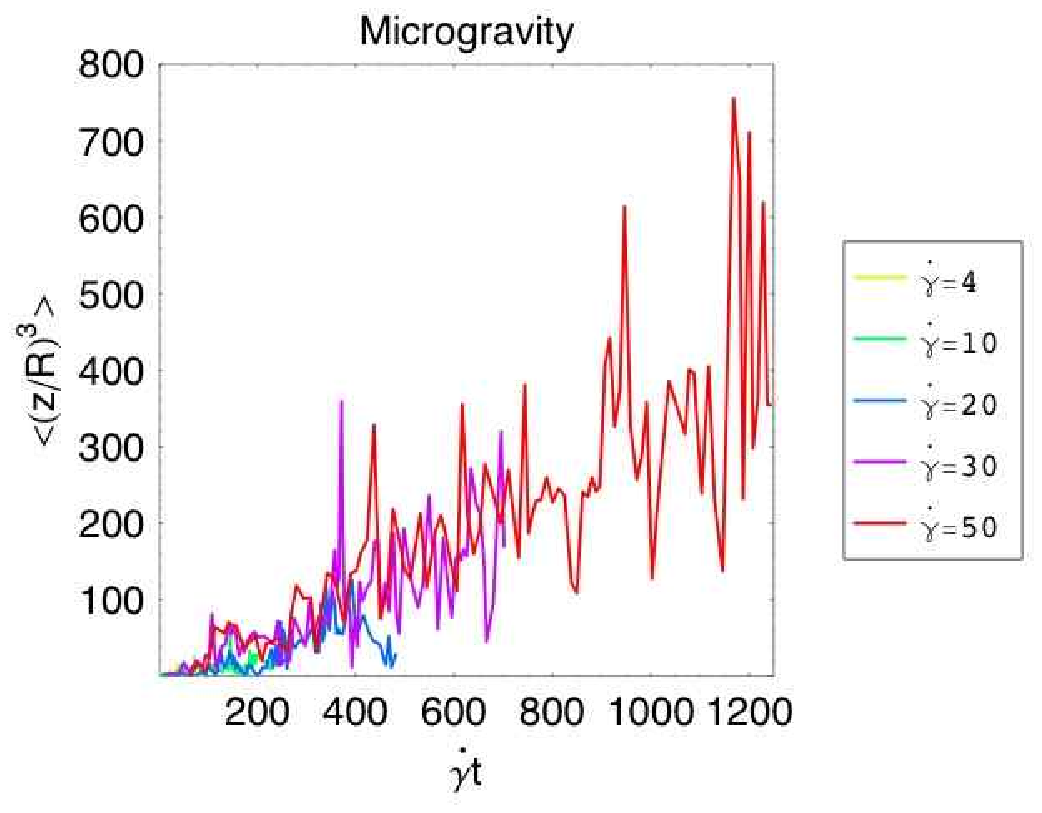} 
\includegraphics[width=8.5cm]{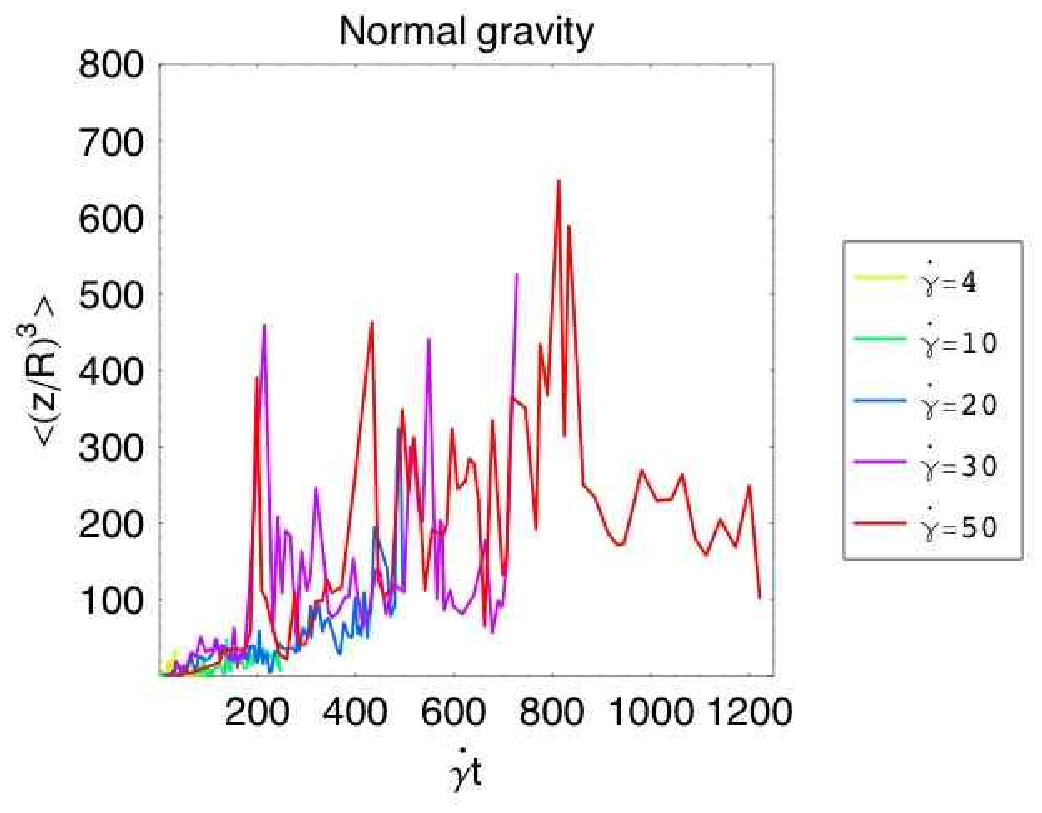} 
\caption{Rescaled vesicle-wall distance $<(z^3/R^3)>$ vs $\dot{\gamma} t$ under microgravity (top) and normal gravity (bottom)conditions for different shear rates (filtering parameters: $5<R<50~\mu$m, $1.1<a_1/a_2<1.25$ ; averaging over 15 detections). \label{fig_resczavg}}
\end{figure}

This suggests to redraw the results with rescaled parameters $(z/R)^3$ and $\dot{\gamma} t$. An example is shown in Fig. \ref{fig_resczavg} where a comparison of results in microgravity and normal gravity is made. Remarkably, in the absence of gravity, all results fall on the same straight line, in agreement with eq. \ref{scaling}, while under gravity a saturation appears, at a value which depends on shear rate. This corresponds to the case studied by Abkarian \emph{et al.}  \cite{Abkarian} where gravity balances the lift force as the latter decreases when the vesicle moves away from the wall. Note that significant scattering is still present on this figure, partly due to statistics and mainly to vesicle dispersion linked to hydrodynamic interactions.

\begin{figure}
\centering
\includegraphics[width=8.5cm]{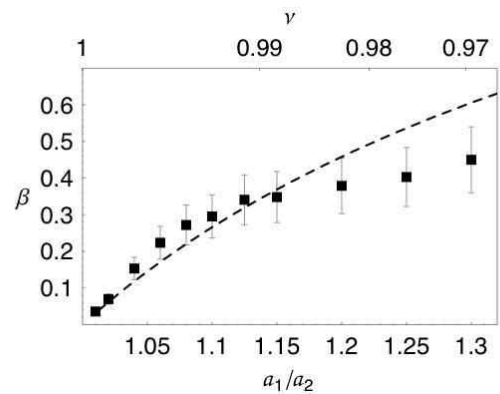} 
\caption{Lift parameter $\beta$ vs vesicle aspect ratio $a_1/a_2$ determined from experimental data. Top horizontal axis: corresponding values of the reduced volume $\nu$. Dashed line: theoretical prediction from Olla  \cite{Olla97} \label{fig_beta}}
\end{figure}

The slope of the data in Fig. \ref{fig_resczavg} should be $\beta=3U$, a function of vesicle aspect ratio and viscosity. To determine the dependence of $\beta$ upon vesicle shape, Fig. \ref{fig_resczavg} was redrawn by varying the range of aspect ratios used for the filtering of data, and a linear fit was performed. One expects vesicles with aspect ratios close to 1 (sphere) to have very small drift velocities (as the lift vanished for a perfect sphere in Stokes flow), while elongated vesicles should drift faster. The results are shown in Fig. \ref{fig_beta} where the slope $\beta$ is plotted vs vesicle aspect ratio, and compared to Olla's prediction (with no fitting parameter). The agreement is quite good both qualitatively and quantitatively in the range of available aspect ratios, except above $1.2$ where the number of corresponding vesicles decreases and statistics are not as favorable.

While Olla's model seems to work quite well for a viscosity ratio of 1 as shown in this paper, it will be interesting in the future to vary this parameter and see how the lift survives when approaching the tank-treading to tumbling transition. The study of the dispersion of vesicles in the z direction should bring interesting information about hydrodynamic interactions under shear.

These different results show that after 20 s of microgravity, no equilibrium distribution seems to be reached, underlining the motivation for longer microgravity times. As far as rheology is concerned, the observed depletion of vesicles near the wall should lead to a decrease of the effective viscosity of the whole sample, as this increasing vesicle-free layer acts as a lubrication layer where viscous dissipation is lower.

\section{Acknowledgements}

The authors wish to acknowledge financial support and access to parabolic flights from the European Space Agency (ESA) and Centre National d'Etudes Spatiales (CNES). This work was supported by the SSTC/ESA-PRODEX (Services Scientifiques Techniques et Culturels/European Space Agency-Programmes de D\'eveloppement d'exp\'eriences) contract 90171. The shear flow chamber used for CNES VP 65 was kindly provided by the Swedish Space Corporation (SSC). We thank V. Vitkova for experimental advice and help, C. Lockowandt (SSC) for assistance during the CNES VP65 campaign, P. Queeckers for help during the ESA PF 43 campaign, G. Giraud, P. Ballet and A. Muylaert for experimental design and technical assistance.

\end{document}